\documentclass[%
 reprint, amsmath,amssymb,aps]{revtex4-2}

\usepackage{graphicx}
\usepackage{dcolumn}
\usepackage{bm}
\usepackage{subcaption}
\usepackage{array}
\usepackage{amssymb}
\usepackage{hyperref}
\hypersetup{
colorlinks=true,
linkcolor=cyan,
filecolor=red,
urlcolor=blue}

\usepackage[font=small,labelfont=bf,justification=RaggedRight]{caption}

\usepackage{subcaption}  

\begin{document}


\title{Search for continuous gravitational waves from HESS~J1427-608\\ with a hidden Markov model}

\author{Deeksha Beniwal$^{1,3}$}
\author{Patrick Clearwater$^{2,4}$}
\author{Liam Dunn$^{2,4}$}
\author{Lucy Strang$^{2,4}$}
\author{Gavin Rowell$^{1,3}$}
\author{Andrew Melatos$^{2,4}$}
\author{David Ottaway$^{1,3}$}
\altaffiliation{}

\email{deeksha.beniwal@adelaide.edu.au}

\affiliation{$^{1}$Department of Physics, University of Adelaide, Adelaide, South Australia 5005, Australia}%
\affiliation{$^{2}$School of Physics, University of Melbourne, Parkville, Victoria 3010, Australia}%
\affiliation{$^{3}$ OzGrav-Adelaide, Australian Research Council Centre of Excellence for Gravitational Wave Discovery, Adelaide, South Australia 5005, Australia}
\affiliation{$^{4}$OzGrav-Melbourne, Australian Research Council  Centre of Excellence for Gravitational Wave Discovery, Parkville, Victoria 3010, Australia}

\date{\today}

\begin{abstract} 
We present a search for continuous gravitational wave signals from an unidentified pulsar potentially powering HESS~J1427-608, a spatially unresolved TeV point source detected by the High Energy Stereoscopic System (HESS). The search uses a semi-coherent algorithm, which combines the maximum likelihood $\mathcal{F}$~statistic with a hidden Markov model to efficiently detect and track quasi-monochromatic signals that wander randomly in frequency. It uses data from the second observing run of the Advanced Laser Interferometer Gravitational-Wave Observatory. Multi-wavelength observations of the HESS source are combined with the proprieties of the population of TeV-bright pulsar wind nebulae to constrain the search parameters. We find no evidence of gravitational wave emission from this target. We set upper limits on the characteristic wave strain $h_0^{95\%}$ (for circularly polarised signals) at $95\%$ confidence level in sample sub-bands and interpolate it to estimate the sensitivity in the full band. We find $h_0^{95\%} = 1.3\times 10^{-25}$ near 185~Hz. The implied constraints on the ellipticity and \textit{r}-mode amplitude reach $\epsilon\leq 10^{-5}$ and $\alpha \leq 10^{-3}$ at 200~Hz, respectively.
\end{abstract}
\maketitle
\section{\label{sec:Introduction}Introduction}
The detection of transient gravitational waves (GWs) from compact binary coalescence events by the Advanced Laser Interferometer Gravitational-Wave Observatory~\cite[aLIGO][]{Aasi_2015} and Advanced Virgo~\cite[aVirgo][]{Acernese_2015} detectors has started a new era of GW astronomy. Steady improvements to the detectors have resulted in more frequent detection of signals from merging binary black holes, binary neutron stars, and neutron star-black hole systems~\cite{Abbott_2019,GWTC2_2021,NSBH_2021}. Isolated, non-axisymmetric neutron stars can also generate persistent, quasi-monochromatic signals detectable by ground-based interferometers \cite{Riles_2013,Riles_2017}. Significant effort has gone into developing tools and methods to search for these continuous gravitational wave (CW) signals. The three main types of CW searches, in order of increasing computational cost, are (1) targeted searches, where the source location and spin parameters are known electromagnetically, e.g.,~\cite{gml_2020, lkw_2020, hvc_2021, quq2021}; (2) directed searches, where only the sky position of the source is known, e.g.,~\cite{mwxAbbott_2021, ozrAbbott_2021, inrAbbott_2021, Millhouse_2020,ScoX1-2022,directed_ulb-2019, cygnusX1-2020}; and (3) all-sky searches where both the location and rotation parameters are unknown, e.g.,~\cite{qhb_2020,una_2021,allsky_ldt-2021,allsky_ulc-2022, semi_ulc-2018,Oliver_2019,Falcon-2019, constraint_ulb-2019}.

In this paper, we perform a directed search for CW signals from a probable neutron star powering a TeV source using public data from the aLIGO second observing run (O2). More specifically, we focus our analysis on HESS~J1427-608, an unidentified and spatially unresolved point source detected by the High Energy Stereoscopic System (HESS) in 2007~\cite{Aharonian_2008}. Using a two-dimensional symmetric Gaussian, the HESS Galactic plane survey estimated a maximum angular diameter of $\sigma=0.048\pm0.009^\circ$ for this source. As the distance to the source is unknown, the small angular extent may indicate either a young or distant source~\cite{Aharonian_2008}. 
Recent analysis of the x-ray multi-mirror mission (XMM-Newton) source catalog found a point-like object, 4XMM J142756.7-605214, located near the center of the HESS emission region~\cite{Devin_2021,Webb_2020}. An extended non-thermal x-ray emission (Suzaku J1427-6051) is also associated with this TeV source~\cite{Fujinaga_2013}. Its x-ray flux is dominated by a central bright source instead of a shell structure typically seen with other supernova remnants (SNRs). The \textit{Fermi} large area telescope (Fermi-LAT) also found a GeV $\gamma$-ray point source, 3FHL~J1427.9-6054, which is spatially coincident with the HESS emission region and has a hard, pulsar-like spectrum that connects smoothly to the TeV $\gamma$-ray spectrum measured by HESS. Although the exact mechanism responsible for the generation of $\gamma$-ray emission is not yet clear, Devin \textit{et al.}~\cite{Devin_2021} conclude that the presence of coincident, extended x-ray emission region and the hard spectral shape of the Fermi-LAT source suggests a leptonic scenario. In this scenario, the TeV $\gamma$-rays are generated by inverse Compton scattering of low-energy photons by relativistic non-thermal electrons or positrons. It is frequently used to describe TeV emission from a pulsar wind nebula (PWN) powered by a young pulsar. The radio morphology and spectral index of the source also indicate a center-filled PWN as opposed to a shell-type SNR~\cite{Abramowski_2011_2}. This, again, provides evidence for the scenario where HESS~J1427-608 is a compact PWN powered by a young pulsar.

The TeV emission could also be a reliable cosignature of CWs as it is associated with young pulsars. These stars are especially likely to be non-axisymmetric, as the mass and current quadrupoles produced during a violent birth have had relatively little time to relax via viscous, tectonic or Ohmic processes~\cite{Chugunov_2010,Knispel_2008,Wette_2010}. Moreover, young pulsars are surrounded by active magnetospheric and particle production processes, which may react back on the star and induce non-axisymmetric variations in the stellar temperature and hence mass density, thus yielding a gravitational wave-emitting mass quadrupole~\cite{Harding_1998,Harding_2002,Levinson_2005,Muslimov_2003}. Therefore, the above characteristics make HESS~J1427-608 an interesting GW target. \\


CW searches for sources like HESS~J1427-608 are usually carried out in a semi-coherent manner, by running a matched filter whose phase tracks the signal coherently within blocks of time but jumps discontinuously from one block to the next, as the signal parameters (e.g. frequency) evolve. There are two reasons for this. First, the CW signals are weaker than compact binary coalescence signals and require integration times of years to be detected. With such long integration times, the number of matched-filter templates (e.g. for the signal frequency and its derivatives in a Taylor expansion) grows prohibitively large, if the search is fully coherent. Second, pulsar timing measurements reveal that the signal frequency evolves unpredictably. Stochastic spin wandering, also known as timing noise, is a widespread phenomenon in pulsars \cite{Hobbs_2010, Ashton_2015,Suvorova_2017}. It is attributed to various mechanisms including changes in the star’s magnetosphere \cite{Lyne_2010}, spin microjumps~\cite{Janssen_2006}, superfluid dynamics in the stellar interior~\cite{Price_2012,Melatos_2014}, and fluctuations in the spin-down torque \cite{Cheng_1987, Urama_2016}. Characterized as a random walk in some combination of rotation phase, frequency, and spin-down rate, timing noise is particularly pronounced in young pulsars with characteristic ages $\lesssim 10$~kyr \cite{Arzoumanian_1994,Hobbs_2010}. Although spin wandering could, in principle, be modeled by including higher-order derivatives in a Taylor expansion of the phase model, the number of derivatives required would make the search computationally infeasible.

We deploy instead a computationally efficient  algorithm to circumvent this issue. The algorithm combines an existing, efficient, and thoroughly tested maximum likelihood detection statistic called the $\mathcal{F}$~statistic \cite{Jaranowski_1998} with a hidden Markov model (HMM)~\cite{Suvorova_2016}. The $\mathcal{F}$~statistic coherently searches for a constant-frequency signal within a block of data, while the HMM tracks the stochastic wandering of the signal frequency from one block to the next~\cite{Quinn_2001}.

The search in this paper uses data from the O2 run of the aLIGO detectors in the 20\textendash200~Hz frequency band. We use the coincident x-ray source 4XMM~J142756.7-60521 to estimate the sky position of the central bright source. Additionally, we choose a coherence time ($T_{\rm{coh}}$) of 7.5~h to balance search sensitivity and computational cost. The motivation behind the choice of search location, frequency range, and $T_{\rm{coh}}$ is outlined in Sec.~\ref{sec:setup}. In Sec.~\ref{sec:method}, we briefly review the search procedure, which is similar to the one used in Refs.~\cite{ScoX1_2017,Jones_2021}, including a description of the signal model, $\mathcal{F}$~statistic, and the HMM implementation. Additionally, we outline the procedure used for selecting a detection threshold and briefly discuss the interferometer data. Above-threshold candidates are passed through vetoes to separate instrumental artifacts from astrophysical signals. The outcome of the analysis is summarized in Sec.~\ref{sec:Results}. We also compute upper limits on the characteristic wave strain and estimate the search sensitivity. Astrophysical implications are discussed in Sec.~\ref{sec:Astro_implications}. Finally, we conclude in Sec.~\ref{sec:Conclusion}. \\ 

\section{\label{sec:setup}Search setup} 
In this section, we outline the motivation behind three key parameter choices: the search location (Sec.~\ref{subsec:search_loc}), frequency band (Sec.~\ref{subsec:freq_band}), and coherence time $T_{\rm{coh}}$ (Sec.~\ref{subsec:coherence_timescale}). The coherence time is defined as the duration of the data interval analysed coherently by the $\mathcal{F}$~statistic.

\subsection{Search location \label{subsec:search_loc}}
The centroid of HESS TeV emission is located at right ascension 14h 27m 52s and declination~$-60^\circ 51' 00''$~(J2000). However, the centroid of a TeV-bright PWN is often offset from the associated pulsar and, as a rule of thumb, the offset gets larger with the age of the system~\cite{Abdalla_2018_2}. This is attributed to a combination of the proper motion of the pulsar, asymmetric evolution of the PWN, or an asymmetric pulsar outflow~\cite{Abdalla_2018_2,Vigelius_2007}. The thermal \footnote{These refer to the soft x rays emitted from polar caps, which are heated by the bombardment of relativistic particles streaming back to the surface from the magnetosphere.} and non-thermal \footnote{This refers to the emission from relativistic charged particles being accelerated in the pulsar magnetosphere} x-ray emission, on the other hand, is directly tied to the density of high-energy electrons in the acceleration region close to the pulsar and is frequently used to detect radio-quiet pulsars \cite{Becker_2009,Prinz_2015,Marelli_2015}. The latest XMM-Newton catalog contains a point-like source, 4XMM~J142756.7-605214, located near the center of the HESS emission region. Although a dedicated study is needed to confirm its relation to the coincident extended x-ray emission (Suzaku~J1427-605), it is likely to be the neutron star powering HESS~J1427-608, as is the case for Cassiopeia~A, Crab, Vela, and other similar SNRs~\cite{Gaensler_2006, Hwang_2004, Weisskopf_2000, Santos_2009}. We, therefore, direct the search at 4XMM~J142756.7-605214 located near the center of HESS~J1427-608.

\subsection{Frequency range\label{subsec:freq_band}}
\begin{figure}[h!]
    \centering
    \includegraphics[trim=0.5cm 0.5cm 0cm 0cm,clip=true, width=0.50\textwidth]{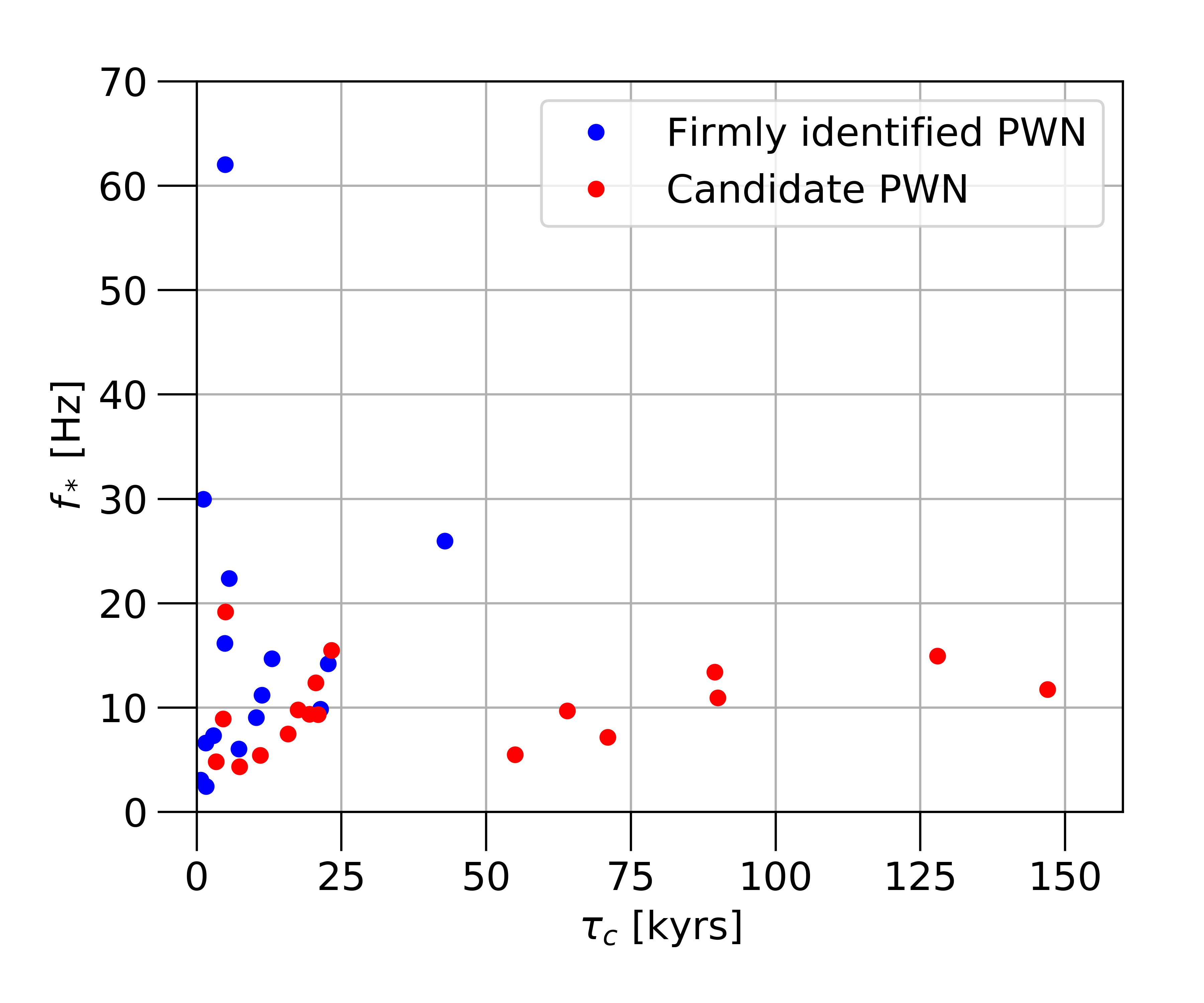}
    \caption{\small{Spin frequency $f_\ast$ versus characteristic age $\tau_c$ of pulsars associated with TeV-bright PWNe in the TeV2 catalog~\cite{Wakely_2013,Camilo_2001,Acero_2013,Manchester_2005}. The blue dots represent pulsars associated with firmly identified (i.e., definitely) TeV-bright PWNe. The red dots show the pulsars associated with HESS candidate PWNe.}}
    \label{fig:PWN_distribution}
\end{figure}

There is no electromagnetic measurement of the spin frequency of the pulsar potentially powering HESS~J1427-608. Therefore, we use the properties of the TeV PWN population to set bounds on the CW signal frequency. The TeV2 catalog~\cite{Wakely_2013} contains 33 sources that are either a candidate PWN or a firmly identified PWN powered by a known pulsar. The characteristic age ($\tau_c$) of these pulsars is plotted against their spin frequency ($f_\ast$) in Fig.~\ref{fig:PWN_distribution} \cite{Camilo_2001,Acero_2013,Manchester_2005}. A majority of these pulsars have $\tau_c \leq25$~kyr and $f_\ast\leq20$~Hz. Plausible emission mechanisms for these pulsars could include a mass quadrupole caused by a thermoelastic or magnetic mountain, which emits a CW signal at $f_\ast$ and $2f_\ast$ \cite{Ushomirsky_2000,Melatos_2005}, \textit{r}-mode, which emit at roughly $4f_\ast/3$~\cite{Owen_1998}, and a current quadrupole produced by non-axisymmetric circulation of neutron star superfluid pinned to the crust, which emits at $f_\ast$ \cite{Jones_2001}. A minimum frequency cutoff of 1~Hz would be ideal as it allows us to explore all of the aforementioned scenarios (i.e., emission at $f_\ast,\ 4f_\ast/3$, and $2f_\ast$) for objects in Fig.~\ref{fig:PWN_distribution}. However, the one-sided amplitude spectral density (ASD) of the detector noise rises rapidly to $\gtrsim5\times10^{-21}\ \rm{Hz}^{-1/2}$ below 20~Hz, which precludes the detection of any plausible CW signal~\cite{Collaboration_2019}. Additionally, the GW open data are aggressively high-pass filtered at 8~Hz to avoid downstream signal processing issues and thus cannot be used for scientific analysis below 10~Hz~\cite{Cahillane_2017}. Therefore, we only search for CW signals above 20~Hz, while acknowledging that only 15 pulsars powering PWNe in the TeV2 catalog have $2f_\ast$ emission above 20~Hz and only seven have $4f_\ast/3$ emission above this limit.\\

We limit the maximum search frequency to 200 Hz for two reasons: 1) The CW emission from isolated young pulsars associated with TeV-bright PWNe lies in the 20\textendash200~Hz range, based on their measured
spin frequencies~\cite{Abdalla_2018_2,Manchester_2005}. 2) Individual recycled millisecond pulsars (MSPs) with typical ages $>10^9$~yr and spin frequencies $f_\ast>100~$Hz are likely to dominate emission above 200~Hz~\cite{Manchester_2005}. While TeV emission has been linked to a population of MSPs in the globular cluster Terzan5~\cite{Abramowski_2011} and, recently, to very extended ($>20$~pc wide) TeV ``halos"~\cite{Hooper_2022}, the observed x-ray and GeV luminosities of HESS~J1427-608 [$L_X(2-10~{ \rm{keV}})>4\times10^{33}$~erg/s and $L_{\rm{GeV}}(>10~\rm{GeV}) > 3\times10^{34}$ erg/s~~\cite{Aharonian_2008}] are orders of magnitude larger than the observed luminosities of known MSPs in the Galactic field (see Figs.~1, 10, and 11 in Refs.~\cite{Lee_2018, Zhao_2022, Caraveo_2014}, respectively). Therefore, HESS~J1427-608 is unlikely to be a system powered by an isolated MSP. This justifies why we can exclude scenarios where the CW emission frequency is likely to exceed 200~Hz.\\

 The search over 20\textendash200~Hz is divided into 2-Hz-wide sub-bands. This ensures that loud non-Gaussian noise artifacts (e.g., lines) are confined to one sub-band and do not affect the whole analysis. Additionally, we overlap the frequency sub-bands by 0.02~Hz to ensure that there is always a sub-band that fully contains a signal, even if the source is rapidly spinning down. The 0.02~Hz overlap is sufficient to cover the maximum plausible amount of spin wandering during a typical observation of $T_{\rm obs}\lesssim234$~days.

\subsection{Coherence time\label{subsec:coherence_timescale}}
The coherence time $T_{\rm{coh}}$ sets the frequency resolution of the search. It is chosen such that a wandering signal frequency remains in one frequency bin during a single coherent step. Here, we rely on the estimated, age-based spin-down rate ($\dot{f}_0$) to choose an optimal $T_{\rm{coh}}$ for the search. It is estimated using \cite{Sun_2018,Jones_2021,Middleton_2020}
\begin{align}
    -\frac{f_0}{(n_{\mathrm{min}}-1)\tau_{\mathrm{age}}} \leq \dot{f_0} \leq -\frac{f_0}{(n_{\mathrm{max}}-1)\tau_{\mathrm{age}}}, \label{eq:fdot_estimate}
\end{align}
where $f_0$ represents the signal frequency, $\tau_{\mathrm{age}}$ is the age of the source (here assumed to be the characteristic age~$\tau_c$), and $n_{\rm{min}}$ and $n_{\mathrm{max}}$ represent the minimum and maximum value of the braking index $n=f_0\ddot{f_0}/\dot{f_0}^2$, respectively. With $\dot{f}_0$ estimated according to Eq. \ref{eq:fdot_estimate}, we set $T_{\rm{coh}} \leq ({2|\dot{f_0}|})^{-1/2}$ \cite{Sun_2018,ScoX1_2017,Beniwal_2021}. A full derivation of this expression is available in Appendix~\ref{app:t_coh}.\\

Multiple groups have attempted to estimate the age of HESS~J1427-608. Assuming that the TeV source is an evolved PWN, Devin \textit{et al.} \cite{Devin_2021} estimate a characteristic age between 4.9 and 13.6~kyr. Fujinaga \textit{et al.} used the correlation between the ratio of $\gamma$-ray to x-ray flux ($F_{\gamma}/F_{X}$) and $\tau_c$ to obtain $\tau_{c}\sim6.4~$kyr \cite{Fujinaga_2012}. However, as reported in Ref.~\cite{Mattana_2009}, the fit that relates $F_{\gamma}/F_{X}$ to $\tau_c$ has an uncertainty factor of $\sim$2.6, which increases the bounds on the estimated age to $2.5< \tau_c < 16.5$~kyr. Additionally, the measured power-law relation between the distance-independent TeV surface brightness and spin-down power yields an age estimate of $1 <\tau_{c}<20$~kyr~\cite{Abdalla_2018_2}. These estimates indicate a relatively young source. Here, we use the age estimate from x-ray observations (i.e. $2.5 < \tau_c < 16.5$~kyr) as the evolution of x-ray emission is strongly correlated with the evolution of the magnetic field, which in turn depends on the morphological evolution of PWN. In contrast, the TeV $\gamma$-ray emission is from the long-lived electrons which trace the time-integrated evolution of the nebula~\cite{Mattana_2009}.\\ 

Now we determine the range of minimum and maximum spin-down rate ($\dot{f_0}$) expected for a target with $f_0\in[20,200]$~Hz and $\tau_{c}\in [2.5,16.5]$~kyr. As the braking index is unknown for this source, we calculate $T_{\rm{coh}}$ for $n\in[2,5,7]$ to cover the extreme plausible values, based on the current observations~\cite{Melatos_1997,Archibald_2016}. The choice of $n=2$ encompasses the broadest range of $\dot{f_0}$ values, while the $n=5$ and $n=7$ cases cover astrophysical scenarios where the phase evolution is purely due to GW emission and \textit{r}-mode oscillations, respectively. Assuming that the frequency evolution caused by timing noise is much smaller than the secular spin-down rate of the pulsar (see Sec.~V B of Ref.~\cite{Beniwal_2021}), we use Eq.~(\ref{eq:fdot_estimate}) to estimate the range of possible $\dot{f}_0$ and thus $T_{\rm{coh}}$. A list of possible values is presented in Table~\ref{tab:fdot_range}. We can cover all of the aforementioned scenarios with $T_{\rm{coh}}=4$~h. However, this $T_{\rm{coh}}$ is not optimal as a large number of data segments, $N_T\gtrsim 1404$, must be incoherently combined, thus degrading the sensitivity of the search which scales $\propto N_{\rm{T}}^{-1/4}$~\cite{Wette_2012}. To circumvent this issue, we choose an intermediate $T_{\rm{coh}} = 7.5$~h, which corresponds to $\dot{f_0} \in [-6.85\times10^{-10},6.85\times10^{-10}]$~Hz/s. This choice of $T_{\rm{coh}}$ allows us to cover the full frequency and age ranges, if the GW emission is due to mass ($n=5$) and current ($n=7$) quadrupoles, and cover most of the interesting parameter space for $n=2$ (i.e., the entire frequency range for $9.5\leq\tau_c\leq16.5$~kyr and part of the spectrum for $\tau_c\leq9.5$~kyr). Signals with $\dot{f_0}$ outside the covered range can also be partially tracked by the HMM, although with lower sensitivity.\\ 

\begin{table}[h!]
    \centering
    \setlength{\tabcolsep}{4pt}
    \renewcommand{\arraystretch}{1.3}
    \caption{Estimated $\dot{f_0}$ and $T_{\rm{coh}}$ in [20,200]~Hz band.}
    \label{tab:fdot_range}
    \begin{tabular}{c l l l}
        \hline
         $n$ & $\tau_{c}$ & $|\dot{f_0}|\ $(Hz s$^{-1})$ & $T_{\rm{coh}}$ (h)  \\ \hline
        2 & 2.5 & [2.54$\times10^{-10}$, 2.54$\times10^{-9}$] & [3.90, 12.33] \\
        2 & 16.5 & [3.84$\times10^{-11}$, 3.84$\times10^{-10}$] & [10.02, 32.68] \\
        5 & 2.5 & [6.34$\times10^{-11}$, 6.34$\times10^{-10}$] & [7.80, 24.66] \\
        5 & 16.5 & [9.61$\times10^{-12}$, 9.61$\times10^{-11}$] & [20.04, 63.36] \\
        7 & 2.5 & [4.23$\times10^{-11}$, 4.23$\times10^{-10}$] & [9.55, 30.21] \\
        7 & 16.5 & [6.41$\times10^{-12}$, 6.41$\times10^{-11}$] & [24.54, 77.60] \\ \hline
    \end{tabular}
\end{table}

\section{\label{sec:method}Search procedure\protect} 
The search presented here is carried out in two steps. First, the $\mathcal{F}$~statistic implementation in \texttt{LALSuite}~\cite{LALapps_2018} is used to coherently combine detector data and compute a log-likelihood score for each 7.5~h time block. Second, HMM tracking is used to find the optimal frequency path through the data over the total observing run. To be precise, the optimal path is recovered using the Viterbi algorithm, a dynamic programming implementation of a HMM. This is similar to the approach used in Refs.~\cite{Suvorova_2016,Suvorova_2017,Sun_2018,Middleton_2020}. The signal model is briefly reviewed in Sec.~\ref{subsec:Sig_model}, while the $\mathcal{F}$~statistic and HMM are outlined in Secs.~\ref{subsec:F_statistic} and \ref{subsec:HMM_framework}, respectively. The procedure used for setting an appropriate threshold is described in Sec.~\ref{subsec:threshold}. Finally, the details of interferometric data used here are outlined in Sec.~\ref{subsec:LIGO_data}.

\subsection{ \label{subsec:Sig_model}Signal model}
We use the signal model described in Ref.~\cite{Jaranowski_1998}. What follows is a summary of the most salient details. We model the wave strain from a neutron star as
\begin{equation}
    h(t) = \mathcal{A}^\mu h_\mu(t), \label{eq:sig_model}
\end{equation}
where $\mathcal{A}_{\mu}$, which depends on characteristic strain amplitude ($h_0$), source orientation ($\iota$), initial phase ($\phi_0$), and wave polarisation ($\times$ or +), represent the amplitudes associated with four linearly independent components~\cite{Jaranowski_1998} 
\begin{align}
    h_{1}(t) &= a(t)\cos\Phi(t),\\ \label{h_start}
    h_{2}(t) &= b(t)\cos\Phi(t), \\
    h_{3}(t) &= a(t)\sin\Phi(t),\\
    \quad h_{4}(t) &= b(t)\sin\Phi(t). \label{h_end}
\end{align}
Here, $a(t)$ and $b(t)$ are the antenna-pattern functions as defined by Eqs. (12) and (13) of Ref.~\cite{Jaranowski_1998} and $\Phi(t)$ is the phase of the CW signal of form 
\begin{equation}
    \Phi(t) = 2\pi f_0[t  + \Phi_m(t;\alpha,\delta)] +  \Phi_s(t;f_0^{(k)},\alpha,\delta). 
    \label{eq:sig_phase}
\end{equation}
The $\Phi_m$ term in the above expression represents the time shift introduced by the diurnal and annual motion of the detector relative to the solar system barycenter, while the $\Phi_s$ is the phase shift that results from intrinsic evolution of the source in its rest frame through its frequency derivatives ($f_0^{(k)} = d^{(k)}f_0/dt^{(k)}$  with $k \geq 1$).\\ 

The intrinsic frequency evolution of a CW signal has two components: a secular term, which can be easily modeled by specifying the frequency derivatives $f_0^{(k)}$, estimated using the procedure outlined in Sec.~\ref{subsec:coherence_timescale}, and a stochastic term, which is often difficult to measure and computationally infeasible to track in a coherent search. To circumvent this issue, we use the HMM outlined in Sec.~\ref{subsec:HMM_framework} to track the stochastic evolution of the signal phase and the Viterbi algorithm to efficiently backtrack and find the optimal pathway in frequency.

\subsection{\label{subsec:F_statistic}$\mathcal{F}$~statistic} 
As in Ref.~\cite{Suvorova_2016}, we use the $\mathcal{F}$~statistic to estimate the likelihood of a CW signal being present in the detector data. The time-dependent output of a single detector is assumed to take the form \cite{Jaranowski_1998} 
\begin{align}
   x(t) &= h(t) + n(t), \label{eq:Detector_data}
\end{align}
where $n(t)$ represents stationary, additive noise and $h(t)$ is the wave strain defined in Eq.~(\ref{eq:sig_model}). We start by defining a normalised log-likelihood of the form \cite{Jaranowski_1998}
\begin{equation}
	\ln\Lambda \equiv (x|h) - \frac{1}{2}(h|h),	\label{log_liklihood}
\end{equation}

where the inner product $(\cdot|\cdot)$ is a sum over single-detector inner products and is given by
\begin{align}
    (x|y) &= \sum_{X=1}^{N_{\rm{Det}}} (x^X|y^X)\\ 
          &\approx {\sum_{X=1}^{N_{\rm{Det}}} \frac{2}{S_{h}^{X}(f)} \int_0^{T_{\rm{obs}}} dt\ x^X(t)y^X(t) }. \label{eq:inner_product} 
\end{align}
Here, $N_{\rm{Det}}$ is the number of detectors and $S_{h}^X(f)$ represents the one-sided power spectral density (PSD) of detector $X$ at frequency $f$~\cite{Prix_2007}.
We maximise $\ln\Lambda$ with respect to the four amplitude parameters $\mathcal{A_\mu}$ to find the optimal set of signal parameters. These parameters, known as the maximum likelihood (ML) estimators, are then used to define the $\mathcal{F}$~statistic,
\begin{align}
    \mathcal{F} &\equiv \ln{\Lambda}_{\mathrm{ML}}\\
    &=  D^{-1}[B (x|h_{1})^2 + A(x|h_{2})^2 -2C(x|h_{1}) (x|h_{2}) \nonumber \\ 
  &\quad  +  B(x|h_{3})^2 +A(x|h_{4})^2 -2C(x|h_{3}) (x|h_{4})] 
\end{align}
with $A = (a|a)$, $B = (b|b)$, $C = (a|b)$, and $D = AB - C^2$. A full derivation can be found in Sec IIIA of Ref.~\cite{Jaranowski_1998}. In the case of white Gaussian noise with no signal, the probability density function (PDF) of the $\mathcal{F}$~statistic takes the form of a central $\chi^2$ distribution with 4 degrees of freedom, $p(2\mathcal{F})=\chi^2(2\mathcal{F};4,0)$. When a signal is present, the PDF has a non-central $\chi^2$ distribution with 4 degrees of freedom, $p(2\mathcal{F})=\chi^2(2\mathcal{F};4,\rho_0^2)$, where the non-centrality parameter is given by
\begin{equation}
{\rho_0}^2 =\frac{K{h_0}^2T_{\rm{coh}}}{S_h(f_0)}.
\end{equation}
Here, $S_h(f)$ is computed as a harmonic sum over the detector-specific $S_h^X(f)$, while the constant $K$ depends on the sky location and orientation of the source~\cite{Jaranowski_1998}.  

\subsection{HMM\label{subsec:HMM_framework}}  %
In this search, we model the stochastic wandering of a CW signal frequency as a Markov chain, whereby the unobservable, hidden state variable $q(t)$ transitions between a set of discrete states $\{q_0,q_1,...,q_{N_Q}\}$ at discrete times $\{t_0,t_1,...,t_{N_T}\}$. Meanwhile, the observable state variable $o(t)$ takes values from the set \{$o_1,...,o_{N_O}$\}. As the Markov chain is memoryless, the state of the system at time $t_{n+1}$ depends only on the state at a previous time step $t_{n}$, and the transition probability matrix is given by
\begin{equation}
A_{q_jq_i} = \mathrm{Pr}[q(t_{n+1}) = q_j|q(t_n) = q_i]. \label{transition_prob} 
\end{equation}
The observable state $o_j$ is related to the hidden state $q_i$ via the emission probability matrix of form
\begin{equation}
L_{o_jq_i} = \mathrm{Pr}[o(t_n) = o_j | q(t_n) = q_i].  
\end{equation}

Finally, the model is completed by specifying the probability of the system occupying each hidden state initially, given by the prior vector of form
\begin{equation}
    \Pi_{q_i} = \mathrm{Pr}[q(t_0) = q_i].
\end{equation}
We then find the most probable sequence of hidden states $Q^*$ given   observable state sequence $O$ by finding $Q$ that maximizes~\cite{Suvorova_2016}
\begin{align}
    \mathrm{Pr}(Q|O) &\propto L_{o(t_{N_T})q(t_{N_T})}A_{q(t_{N_T})q(t_{N_{T}-1})}... \nonumber\\
    &\quad \times L_{o(t_1)_q(t_1)}A_{q(t_1)}\Pi_q(t_0).\label{Markov_chain}
\end{align}
The Viterbi algorithm, as described in Ref.~\cite{viterbi_original} and applied to CW searches in Refs.~\cite{Suvorova_2016, Suvorova_2017,Middleton_2020,Millhouse_2020}, provides a recursive and computationally efficient method for computing $Q^*$ from Eqs.~(\ref{transition_prob})\textendash(\ref{Markov_chain}). For computational convenience and numerical stability, we evaluate $\mathcal{L}=\log \mathrm{Pr}(Q|O)$, whereby the products in Eq.~(\ref{Markov_chain}) become a sum of log-likelihoods.\\

Here, the CW signal frequency $f_0(t)$ is the hidden state variable, which moves by at most one bin up or down during a timescale $T_{\rm{coh}}$. This is identical to the approach used in Refs.~\cite{Suvorova_2016,Suvorova_2017,Millhouse_2020} and represented by the following transition matrix: 
\begin{equation}
  A_{q_{i-1}q_i} = A_{q_{i}q_i}  = A_{q_{i+1}q_i} = \frac{1}{3}, \label{eq:TM_old}
\end{equation} 
with all other $A_{q_{j}q_i}$ entries being zero. This choice of matrix is appropriate because timing noise is especially pronounced in young pulsars with $\tau_c \lesssim 10$~kyr \cite{Arzoumanian_1994,Hobbs_2010}. The observable is the $\mathcal{F}$~statistic with an emission probability given by~\cite{Suvorova_2016,Sun_2018}
\begin{equation}
    L_{o_jq_i} \propto \exp{[\mathcal{F}(f_0)]},
\end{equation}
where $\mathcal{F}(f_0)$ is computed for each segment of length $T_{\rm{coh}}$ at a frequency resolution of $\Delta f_{\rm{coh}} = 1/(2T_{\rm{coh}})$. The method for setting $T_{\rm{coh}}$ is discussed in Sec.~\ref{subsec:coherence_timescale}. We choose a uniform prior $\Pi_{q_i}=N_Q^{-1}$ for this study, as the frequency of the signal at $t_0$ is unknown. 

\subsection{Threshold \label{subsec:threshold}}
\begin{figure}[h!]
    \centering
    \includegraphics[scale=0.5, trim=0.45cm 0.45cm 0cm 0cm, clip]{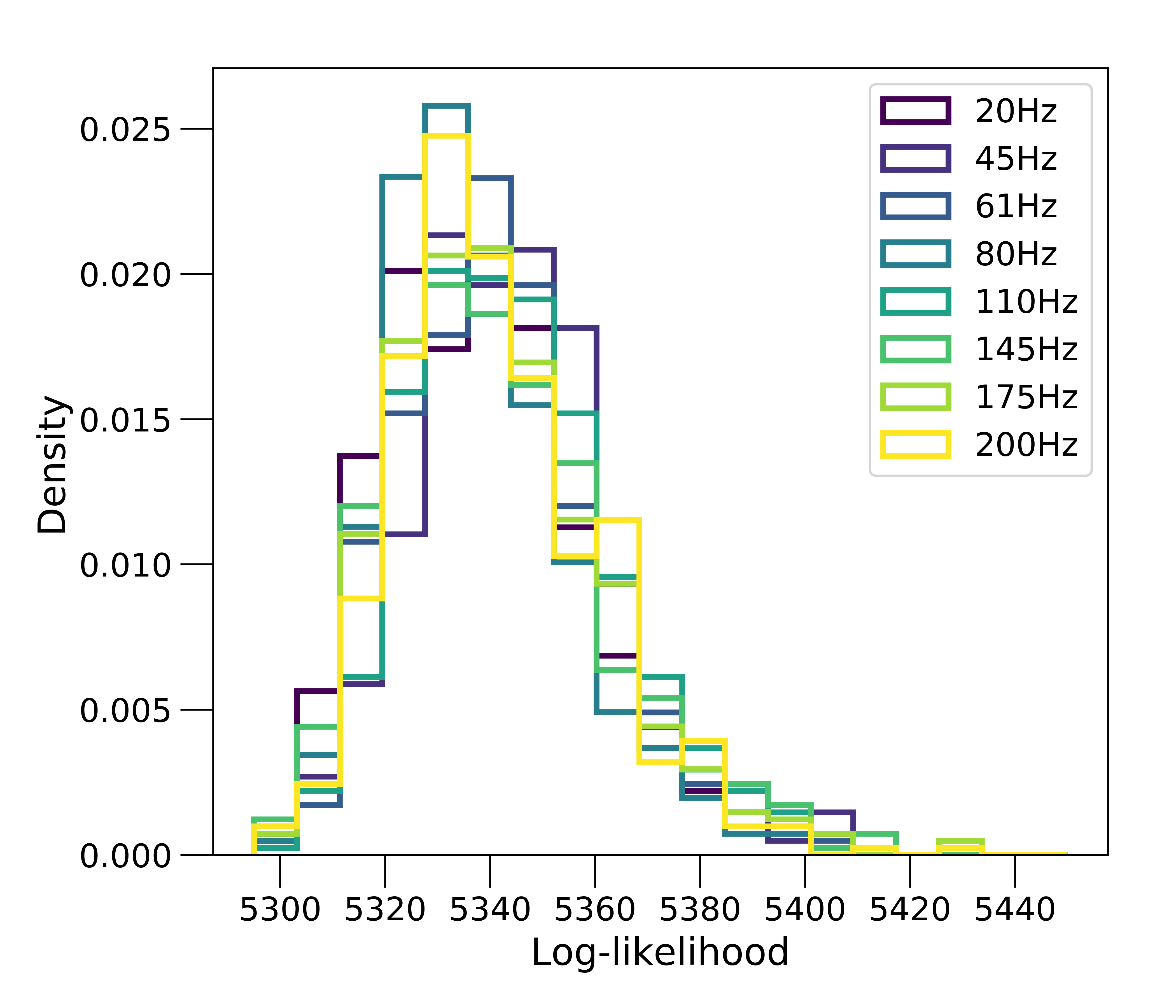}
    \caption{Distribution of maximum log-likelihood scores in Gaussian noise across eight frequency sub-bands. Each distribution is constructed using 500 realisations. The starting frequency of each 2-Hz-wide sub-band is denoted in the legend.}
    \label{fig:threshold_dist}
\end{figure}

We aim to find a threshold log-likelihood ($\mathcal{L}_{\rm{th}}$) that corresponds to a desired false alarm probability $\alpha_f = 1\%$ in each sub-band. The detector ASD $\left[{S_{h}(f)}^{1/2}\right]$ changes by 3 orders of magnitude over the 20\textendash200~Hz band~\cite{Collaboration_2019}. To understand how this may affect the threshold for the search, we generate 500 Gaussian noise realizations in eight, 2-Hz-wide sub-bands, namely, those starting at 20, 45, 61, 80, 110, 145, 175, and 200~Hz. We set ${S_{h}(f)}^{1/2}$ of each sub-band to match the band-averaged one-sided noise ASD of the O2 data~\cite{Collaboration_2019}. The search parameters are outlined in Table \ref{tab:Search_parameters}. Figure~\ref{fig:threshold_dist} shows the distributions of maximum log-likelihoods obtained using 500 realizations in each sub-band, all of which are consistent with a Gumbel distribution~\cite{Tenorio_2022}. We list the 99th percentile $\mathcal{L}_{\rm{th}}$ corresponding to $\alpha_f = 1\%$ for each sub-band in Table \ref{tab:predicted_threshold}. The results show relatively small variation ($\sim 0.2\%$) over the 20\textendash200~Hz frequency range. This is consistent with the results presented in Ref.~\cite{ScoX1_2019}, where a similar trend is observed despite using a slightly different detection statistic. These results also show negligible variation in the predicted $\mathcal{L}_{\rm{th}}$ with the detector noise PSD. This is expected as the inner product defined in Eq.~(\ref{eq:inner_product}) includes a noise weighting $\propto S_h(f)^{-1}$ that removes the dependence on $S_{h}(f)$. We therefore combine all 4000 noise realisations into a single dataset and use it to compute the 99th percentile $\mathcal{L_{\rm{th}}}$ = 5396, corresponding to $\alpha_f = 1\%$ per 2~Hz sub-band.     
    
\begin{table}[h!]
    \centering
    \setlength{\tabcolsep}{4pt}
    \renewcommand{\arraystretch}{1.2}
    \caption{Search parameters. Here, $N_T = T_{\rm{obs}}/T_{\rm{coh}}$ is the total number of coherent steps. RA, right ascension: DEC, declination.}\label{tab:Search_parameters}
    \begin{tabular}{c c c}
\hline    Parameter & Value & Units \\ \hline 
          RA & 14:27:56.7 & J2000 h:min:s\\ 
          DEC & \textendash60:52:14 & J2000 deg:min:s  \\ 
         $f_0$ & 20\textendash200 &~Hz  \\
        $\Delta f_{\rm{coh}}$ &  $1.9\times 10^{-5}$ &~Hz\\
         $T_{\rm{coh}}$ & 7.5 & h\\
         $T_{\rm{obs}}$ & 234 & days\\
        $N_{\rm{T}}$ & 748 & ... \\ \hline
    \end{tabular}
\end{table}

\begin{table}[h!]
    \setlength{\tabcolsep}{5pt}
    \renewcommand{\arraystretch}{1.2}
    \centering
    \caption{The 99th percentile log-likelihood thresholds for eight sub-bands used in the threshold study.}
    \begin{tabular}{c c}
    \hline
        sub-band (Hz]) & $\mathcal{L}_{\rm{th}}$  \\ \hline
         20\textendash22 & 5394 \\
         45\textendash47 & 5402 \\
         61\textendash63 & 5394 \\
         80\textendash82 & 5388 \\
         110\textendash112 & 5399 \\
         145\textendash147 & 5399\\
         175\textendash177 & 5400 \\
         200\textendash202 & 5393 \\ \hline
    \end{tabular}
    \label{tab:predicted_threshold}
\end{table}
\subsection{LIGO data \label{subsec:LIGO_data}}
We use the publicly available data from the O2 run of the aLIGO detectors, specifically the GWOSC-4KHz$\_$R1$\_$STRAIN channel, to search for CW signals from HESS~J1427-608~\cite{Collaboration_2019}. The aVirgo detector was also operating for the last month of O2. However, we omit Virgo data due to the relatively short observing time and lower sensitivity~\cite{Collaboration_2019}. We also exclude data from the first few weeks of the O2 run (November 30, 2016\textendash January 4, 2017), where the data quality was not optimal and was followed by a brief break in the operation of both detectors. With these cuts, we choose a common period from January 4\textendash August 25, 2017 (GPS time = 1167545066\textendash1187762666) to perform a joint analysis of the two aLIGO detectors. This gives us a total observation period of $T_{\rm{obs}}=234$ days. Table \ref{tab:Search_parameters} summarises the parameters of the search.

\section{Results \label{sec:Results}}

\begin{figure}[h!]
    \centering
    \includegraphics[width=0.5\textwidth,trim=0.5cm 0.6cm 0cm 0.5cm,clip=true]{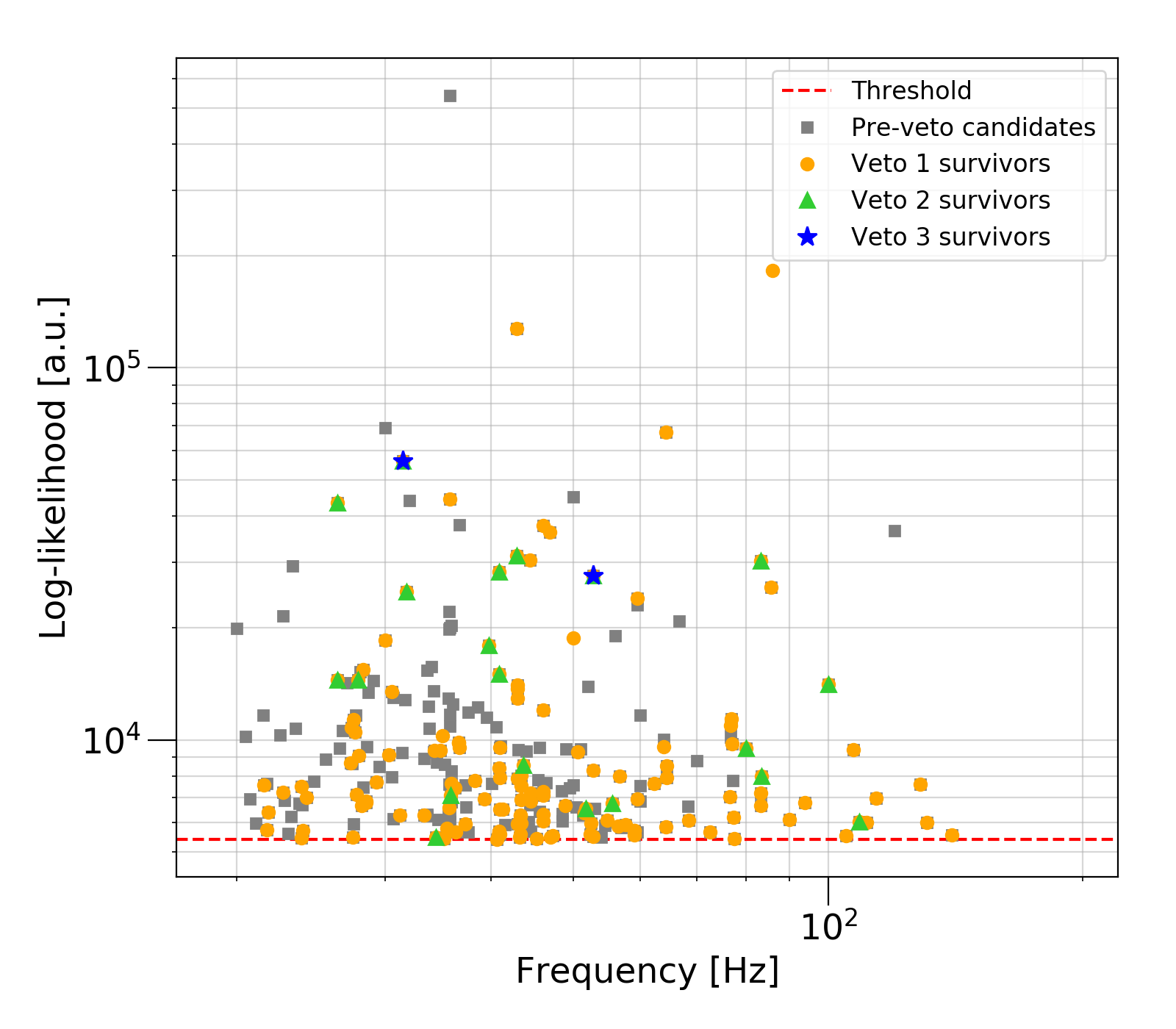}
    \caption{\small{Log-likelihood ($\mathcal{L}$) versus frequency of candidates with $\mathcal{L}>\mathcal{L}_{\rm{th}}$. The dotted horizontal line indicates the Gaussian threshold set using the procedure outlined in Sec.~\ref{subsec:threshold}. The pre-veto candidates (gray squares) are shown along with the survivors after the known lines veto (orange circles), single interferometer veto (green triangles) and Doppler modulation veto (blue stars). No candidate survives the off-target veto.}}
    \label{fig:search_results}
\end{figure}

The search returns a total of 246 candidates with $\mathcal{L}>\mathcal{L}_{\rm{th}}$ over the frequency range from 20\textendash200~Hz. This number exceeds the single candidate expected given $\alpha_f=1\%$ per sub-band, most likely due to non-Gaussian features in the real aLIGO detector noise. This is especially true for sub-bands with $f_0<80$~Hz, which are heavily contaminated by instrumental lines and contain 94$\%$ of candidates identified here (refer to Fig.~\ref{fig:search_results}).\\

All 246 candidates are passed through a four-step veto procedure, which is adapted from previous studies \cite{ScoX1_2017,ScoX1_2019,Jones_2021} and used to eliminate candidates resulting from noise artifacts. The vetoes are as follows: (1)~The known line veto rejects candidates that have a Viterbi path that intersects a known instrumental line in either the Hanford or Livingston detector. (2)~The single interferometer veto eliminates candidates that return $\mathcal{L}>\mathcal{L}_{\cup}$ in only one of the interferometers, not both. Here, $\mathcal{L}_{\cup}$ denotes the log-likelihood from the dual-interferometer search. (3)~The Doppler modulation (DM) veto involves turning off the DM correction, which accounts for the Doppler shift due to Earth's motion, and recomputing the log-likelihoods. Astrophysical signals usually become undetectable without this correction~\cite{Jones_2022}. Therefore, candidates with comparable $\mathcal{L}$ in both the DM-on and DM-off searches are rejected. (4)~The off-target veto involves performing the search at an offset sky location. We reject candidates that return comparable $\mathcal{L}$ in both the on-target and off-target searches.

The reader is referred to Appendix~\ref{app:vetoes} for a detailed description of all four vetoes. Table \ref{tab:veto_summary} summarises the outcome at each veto step, while Fig.~\ref{fig:search_results} shows the location of each candidate in the log-likelihood ($\mathcal{L}$) versus frequency space. No candidates survive at the end of the veto procedure. \\

\begin{table}[h!]
    \centering
    \caption{\small{The number of candidates remaining after each post-processing step. A total of 246 candidates, with $\mathcal{L}>\mathcal{L}_{\rm{th}}$, are passed through a four-step veto procedure. None survive this procedure.}}
    \label{tab:veto_summary}
    \setlength{\tabcolsep}{4.5pt}
    \renewcommand{\arraystretch}{1.2}
    \begin{tabular}{l l}
    \hline 
        Processing step & Number of candidates \\ \hline
        Preveto & 246 \\
        Known lines veto & 135\\
        Single interferometer veto & 20\\
        Doppler modulation veto & 2\\
        Off-target veto & 0 \\ \hline
    \end{tabular}
\end{table}

\subsection{Observational upper limits\label{subsec:Upper_lims}}
No above-threshold candidates survive the postprocessing analysis outlined in Sec.~\ref{sec:Results}. We therefore start by obtaining an estimate for the minimum detectable strain using the following analytic expression for the $95\%$ confidence sensitivity of a  search~\cite{Sun_2018,Wette_2008}, viz.
\begin{equation}
        h_0^{\rm{est}} = \Theta S_h(f)^{1/2}(T_{\rm{obs}}T_{\rm{coh}})^{-1/4}, \label{eq:theoretical_minimum}
\end{equation}
\noindent where $S_h(f)^{1/2}$ is the noise ASD and $\Theta$ is the statistical threshold that is directly proportional to the root-mean-square signal-to-noise ratio of the search~\cite{Wette_2012}. One typically finds $30 \lesssim \Theta \lesssim 40$ for this type of search~\cite{Wette_2008}. Following previous searches for CW signals with a HMM~\cite{Strang_2021,Sun_2018,Wette_2008}, we set $\Theta$ = 35. The theoretical upper limit ($h_0^{\rm{est}}$) is shown by the red curve in Fig~\ref{fig:upper_limits_results}.\\

\begin{figure}[h!]
    \centering
    \includegraphics[width=0.5\textwidth, trim=0.5cm 0.6cm 0cm 0.5cm, clip=true]{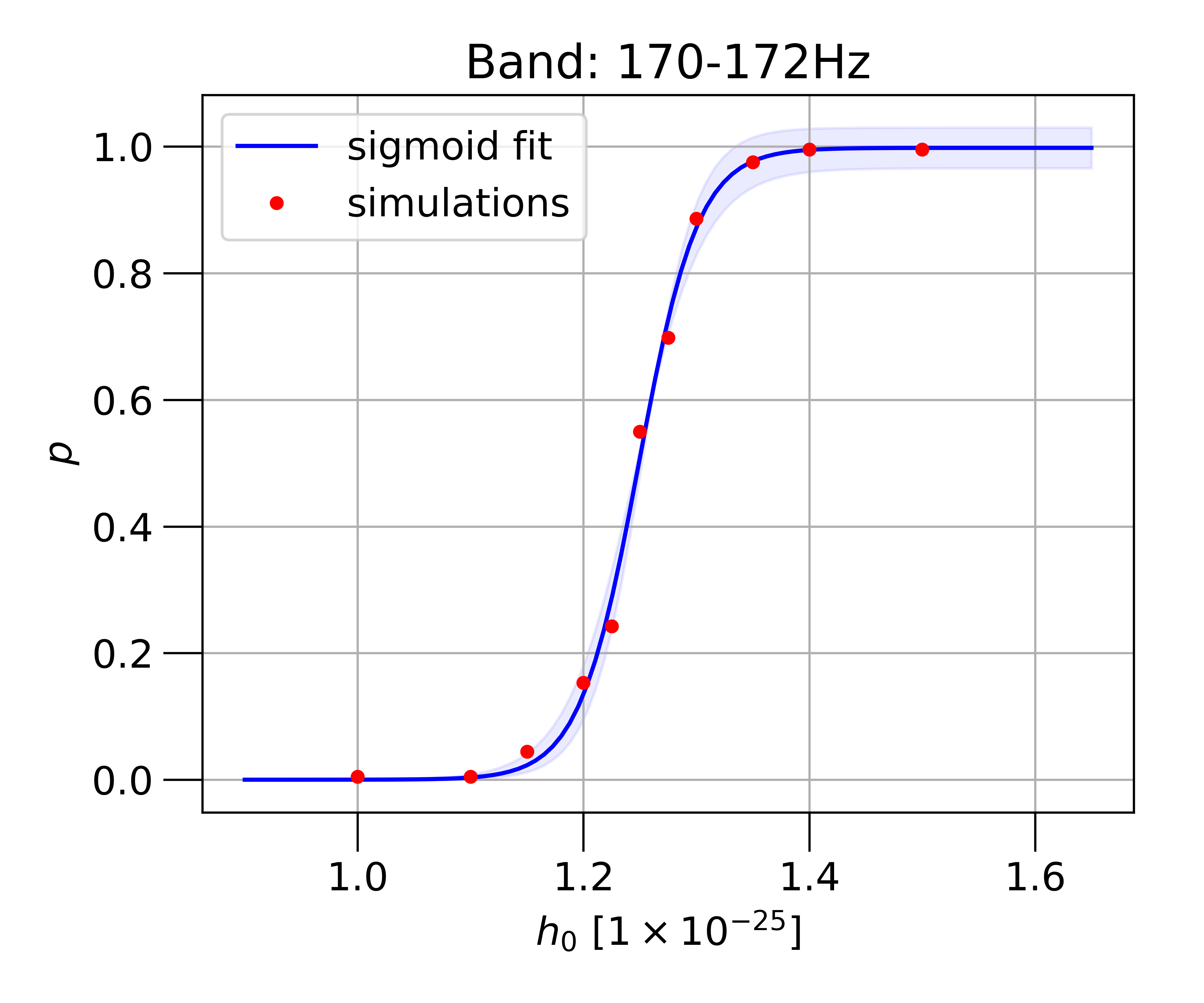}
    \caption{\small{Detection efficiency ($p$) versus signal strain ($h_0$) for 170$-$172~Hz sub-band. The red dots show the results obtained with simulated signals in detector data while the solid blue curve shows the sigmoid fit along with the $2\sigma$ uncertainty interval.} }
    \label{fig:sigmoid_fit}
\end{figure}

We also quantify the sensitivity of the search by estimating $h_0^{95\%}$, such that a circularly polarised signal with $h_0 \geq h_0^{95\%}$ is detectable on 95$\%$ or more occasions. To do this, we randomly inject 100 simulated signals with a fixed $h_0$ into five sub-bands, namely, those starting at 86, 101, 146, 170, and 194~Hz. These sub-bands are chosen at random from a set of bands that return less than three unique paths with $\mathcal{L}> \mathcal{L_{\rm{th}}}$ in the original search. Equation~(\ref{eq:theoretical_minimum}) provides the scaling that can be used to estimate the sensitivity of all other frequency sub-bands. Additionally, the frequencies of these software injections are randomly chosen within the sub-band. As the upper limits computed here are model dependent, we set $\cos \iota = 1$ for optimal orientation and randomly draw the polarisation angle from $\psi \in [0, \pi]$. 
For each trial, we then use a combination of the $\mathcal{F}$~statistic and Viterbi algorithm to look for the injected signal in the detector data, using the setup outlined in Table~\ref{tab:Search_parameters}. The above process is repeated for ten different values of $h_0$ in each sub-band. Each trial acts as a Bernoulli trial with a probability of success (efficiency) $p$ given by the Wilson interval~\cite{Wilson_1927},
\begin{multline}
    p \approx \frac{s+\frac{1}{2}{(1-\alpha_f/2)}^2}{N_I + {(1-\alpha_f/2)}^2}\pm \frac{1-\alpha_f/2}{N_I + {(1-\alpha_f/2)}^2} \\
    \sqrt{\frac{s(N_I-s)}{N_I}+\frac{{(1-\alpha_f/2)}^2}{4}},
\end{multline}
where $N_I = 100$ is the number of injections and $s$ is the number of successes. For each sub-band, we use the curve fit tool in \texttt{PYTHON} to fit a sigmoid curve~\cite{Banagiri_2019} to the distribution of recovery efficiency ($p$) versus strain sensitivity ($h_0$), with uniform priors over the sigmoid parameters. An example fit is shown in Fig.~\ref{fig:sigmoid_fit}. The best fit parameters are then used to find $h_0^{95\%}$ which corresponds to $p=95\%$. The black dots in Fig.~\ref{fig:upper_limits_results} show the estimated upper limits for the five sub-bands tested here.  \\ 

Finally, we calculate $a=h_0^{95\%}/h_0^{\rm{est}}$ for each of the five sub-bands tested here and use the mean $a$ across these sub-bands to estimate the sensitivity across the full frequency band as $h_0^{95\%} = ah_0^{\rm{est}}$. This is shown by the blue curve in Fig.~\ref{fig:upper_limits_results}.

The estimated search sensitivity is below the theoretical limit over the entire search band. This occurs because the search is sensitive not to $h_0$ but to a combination of $h_0$ and $\cos{\iota}$, commonly referred to as $h_0^{\rm{eff}}$. The scaling is given by \cite{Messenger_2015}
\begin{equation}
    \left(h_0^{\rm{eff}}\right)^2 = h_0^2 \frac{[(1 + \cos^2
\iota)/2]^2 + [\cos\iota]^2}{2},
\end{equation}
where $(h_0^{\rm{eff}})^2 = h_0^2$ for circular polarisation (i.e.,~$\iota = 0^\circ$~or~$180^\circ$), $h_0^2/8$ for linear polarisation (i.e., $\iota=90^\circ$) and $2h_0^2/5$ for an isotropic average over the inclination angle $\iota$.\\

The $h_0^{95\%}$ limits obtained empirically in the five sample sub-bands assume $\cos{\iota} = 1$ for optimal orientation (i.e., circularly polarised signals). Therefore, these limits do not need to be scaled as one has $h_0^{\rm{eff},95\%} = h_0^{95\%}$. The expression for $h_0^{\rm{est}}$, on the other hand, assumes marginalisation over the unknown parameters, namely $\cos{\iota}$ and $\psi$~\cite{Wette_thesis2009, Wette_2012}. Therefore, one must scale $h_0^{\rm{est}}$ to obtain an effective sensitivity independent of $\cos \iota$ (i.e., $h_0^{\rm{eff,est}} =\sqrt{2/5}\ h_0^{\rm{est}}$) before comparing it to $h_0^{\rm{eff,95\%}}$ or $h_0^{\rm{95\%}}$.\\

The scaling ratio, which is used to compute sensitivity across the full frequency band, is thus expected to be $a = h_0^{\rm{eff, 95\%}}/h_0^{\rm{eff, est}} = 0.633$. Based on the simulation results, the calculated ratio (i.e., $a = 0.639$) is consistent with this value within uncertainty and thus explains the observed trend.

\subsection{Theoretical upper limits}
\begin{figure}
    \centering
    \includegraphics[width=0.5\textwidth,  trim=0.55cm 0.55cm 0cm 0.6cm, clip=true]{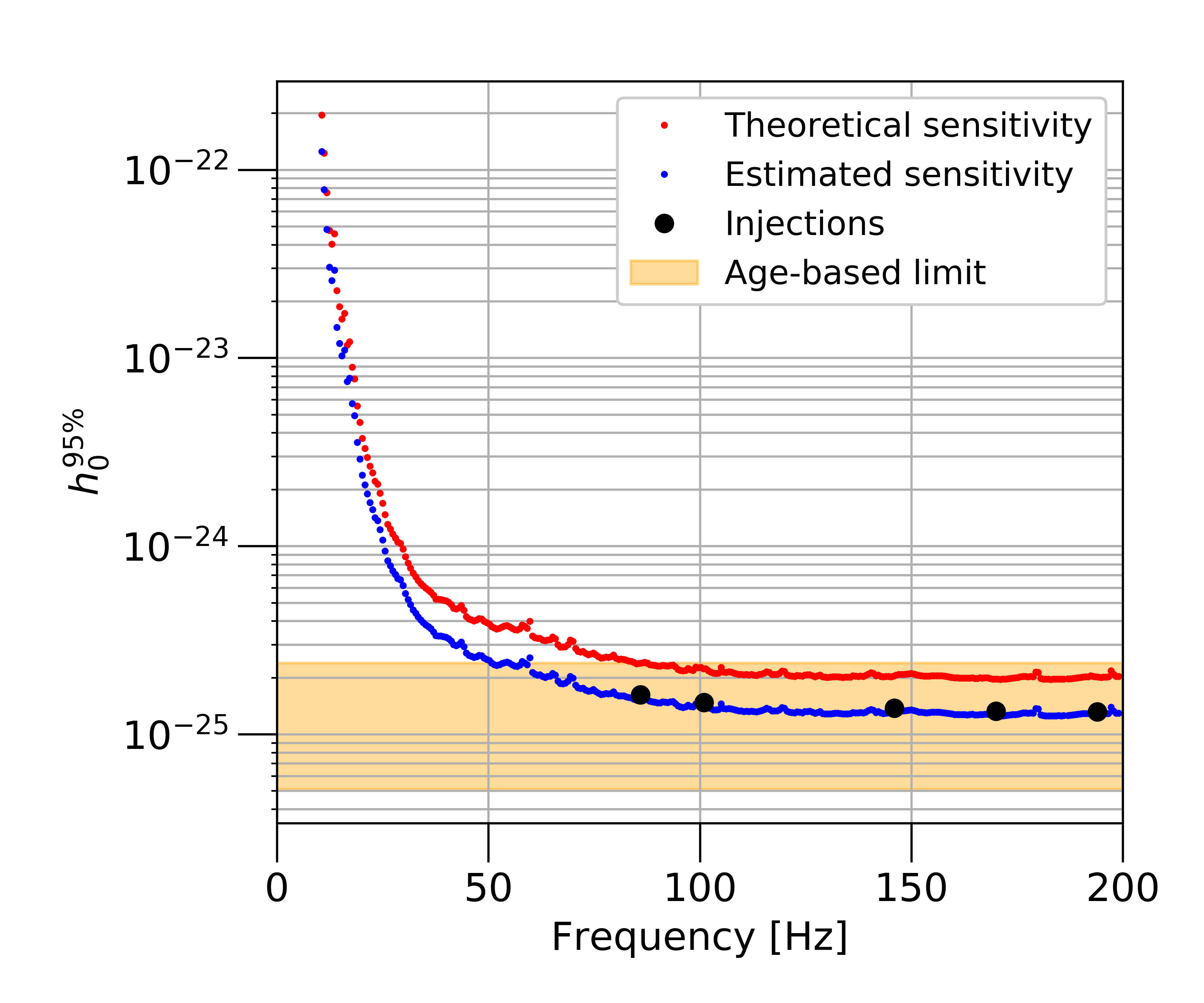}
     \caption{\small{The sensitivity estimate $h_0^{95\%}$ obtained for HESS~J1427-608 assuming a circularly-polarised signal. The red curve represents the minimum detectable strain derived using the analytic expression in Eq.~(\ref{eq:theoretical_minimum}). The black dots represent the $h_0^{95\%}$ limit obtained empirically in the sample sub-bands. The blue curve represents the estimated $h_0^{95\%}$ across the full 20\textendash200~Hz band. The orange band indicates the age-based indirect limit ($h_0^{\rm{age}}$).}}
    \label{fig:upper_limits_results}
\end{figure}

Assuming that the star loses all of its rotational kinetic energy as GWs, we can also derive an age-based indirect strain limit ($h_0^{\rm{age}}$) for the source, given by \cite{Strang_2021}
\begin{equation}
    h_0^{\rm{age}} = 2.27\times 10^{-24}\left(\frac{\rm{1~kpc}}{D}\right)\left(\frac{\rm{1~kyr}}{\tau_{c}}\right)^{\frac{1}{2}}\left(\frac{I_{zz}}{\rm{10^{38}~kg~m^2}}\right).
\end{equation}
Here, $D$ is the distance to the source, $\tau_{c}$ is the characteristic age, and $I_{zz}$ is the principle moment of inertia. Although the exact distance to HESS~J1427-608 is unknown, studies suggest a value between 6 and 11~kpc \cite{Fujinaga_2012,Venter_2018,Devin_2021}. Assuming $I_{zz}=10^{38}$ kg m$^2$ and $\tau_c~\in~[2.5, 16.5]$~kyr, the age-based indirect strain sensitivity is $0.51 \leq h_0^{\rm{age}} / (10^{-25}) \leq 2.39$. This is shown by the horizontal orange band in Fig.~\ref{fig:upper_limits_results}. Note that the 95$\%$ confidence upper limit surpasses the indirect age-based limit for $f_0>50$~Hz and specific source scenarios (i.e., $2.5< \tau_c<4$~kyr and $6< D <7.5$~kpc), reaching a minimum of $1.3\times10^{-25}$ around 185~Hz. Data from O3 and future observing runs are needed to explore the remaining parameter space and exclude other source scenarios.

\section{Astrophysical implications}\label{sec:Astro_implications}
\begin{figure}
    \centering
    \begin{subfigure}{0.5\textwidth}
        \caption{Ellipticity upper limit}
        \includegraphics[trim=0.4cm 0cm 0.5cm 0.9cm, clip, width=1\textwidth]{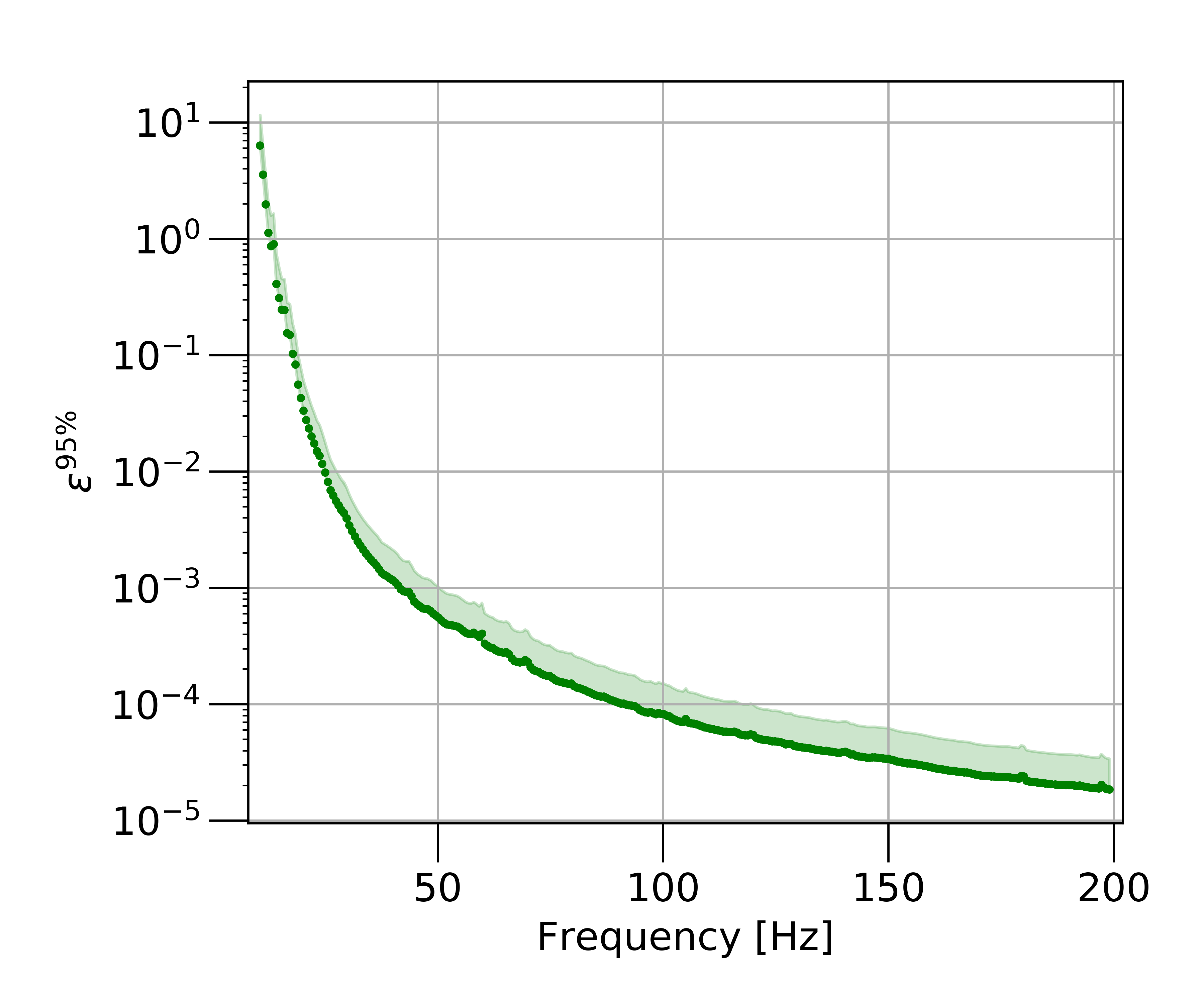}
    \end{subfigure}
    
    \begin{subfigure}{0.5\textwidth}
        \caption{R-mode amplitude upper limit}
        \includegraphics[trim=0.4cm 0.5cm 0.5cm 0.9cm, clip, width=1\textwidth]{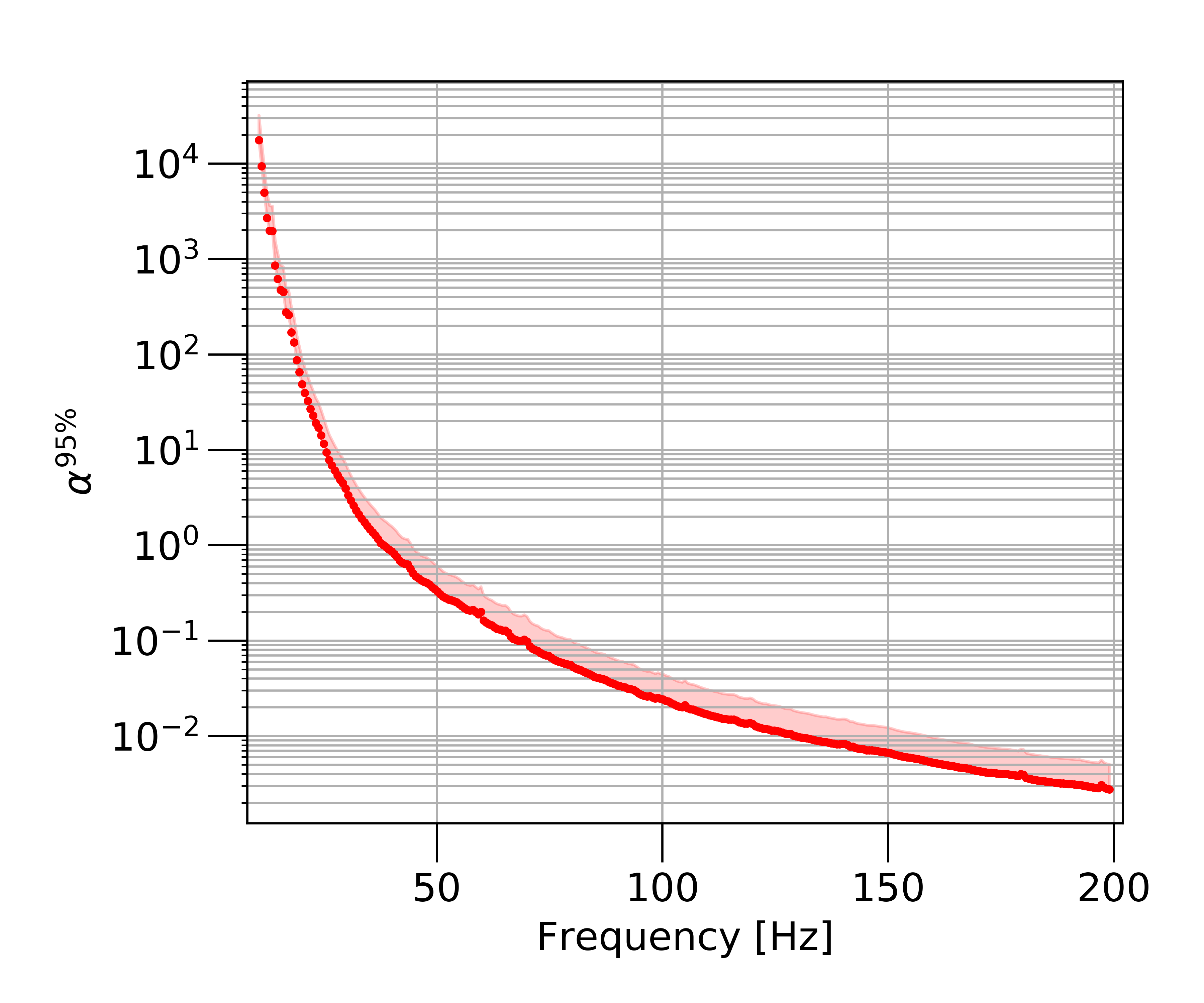}
    \end{subfigure}
    \caption{\small{Constraints on the (a) neutron star ellipticity and (b) \textit{r}-mode amplitude, plotted as a function of CW signal frequency ($f_\ast$). The red and green dotted lines correspond to the closest distance estimate (i.e., $D=6$ kpc), while the shaded regions indicate the results across the full distance range.}}
    \label{fig:epislon_alpha}
\end{figure}

The GW strain upper limit can be converted into a constraint on the ellipticity of the neutron star $\epsilon$~\cite{Jaranowski_1998} and the \textit{r}-mode amplitude parameter $\alpha$~\cite{Owen_1998}. For ellipticity calculations, we assume that the CW signal frequency ($f_0$) is twice the rotational frequency ($2f_\ast$). Given the estimated $h_0^{95\%}$ and assuming a canonical moment of inertia ($I_{zz} = 10^{38}$ kg m$^2$), we constrain the fiducial ellipticity of the neutron star in terms of the GW frequency via \cite{Jaranowski_1998, Wette_2008}
\begin{equation}
    \epsilon = 9.46\times 10^{-5} \left(\frac{h_0^{95\%}}{10^{-24}}\right)  \left(\frac{D}{\rm{1~kpc}}\right)\left(\frac{\rm{100~Hz}}{f_0}\right)^2.
\end{equation}
We also convert $h_0^{95\%}$ to a limit on the amplitude of \textit{r}-mode oscillations via \cite{Owen_2010, Lindblom_1998}
\begin{equation}
    \alpha \simeq 0.028 \left(\frac{h_0^{95\%}}{10^{-24}}\right)  \left(\frac{D}{\rm{1~kpc}}\right)\left(\frac{\rm{100~Hz}}{f_0}\right)^3.   
\end{equation}
These limits are shown in Fig.~\ref{fig:epislon_alpha}. The best constraints on the star's ellipticity ($\epsilon \leq 2\times 10^{-5}$) and \textit{r}-mode amplitude ($\alpha \leq 3\times10^{-3}$) are obtained at 200~Hz. The ellipticity constraint is above the rough theoretical maximum ($\epsilon \sim 10^{-6}$) predicted for a neutron star~\cite{Johnson_2013, Baiko_2018, Gittins_2021}. The constraint on the \textit{r}-mode amplitude, however, does reach the $\alpha \sim 10^{-3}$ level expected for the most detailed exploration of the nonlinear saturation mechanism \cite{Bondarescu_2009, Haskell_2015}. Data from future observing runs should provide stricter, more meaningful constraints. Note that the results presented here refer to a specific scenario with source properties $\psi \in [0,\pi]$ and $\cos \iota = 1$. 

\section{Conclusion} \label{sec:Conclusion}
In this paper, we present a search for CW signals from HESS~J1427-608, an unidentified and spatially unresolved TeV point source potentially harboring a young pulsar, using aLIGO's publicly available O2 data. The search uses a method that combines the maximum likelihood $\mathcal{F}$~statistic with a HMM to efficiently track the secular spin-down and stochastic spin wandering. No evidence of a CW signal is found. We set the first upper limits on the GW signal amplitude, mass ellipticity ($\epsilon$), and \textit{r}-mode amplitude ($\alpha$) for this target in a directed search. The best (lowest) constraint on the CW signal amplitude (for circularly polarized signals) is $h_0^{95\%} \approx 1.3 \times 10^{-25}$ near 185~Hz, while $\epsilon$ and $\alpha$ are constrained to $< 2\times 10^{-5}$ and $<3\times 10^{-3}$ around 200~Hz, respectively.\\ 

HESS~J1427-608 has not been targeted by previous CW searches. Therefore, we cannot make a direct comparison between the results presented here and previous studies. We can, however, qualitatively compare these results with searches that target young SNRs with a compact central object ($\tau_{c}\in [2.5, 16.5]$~kyr) and use O2 data. Lindblom and Owen~\cite{Lindblom_2020} used O2 data to search for signals from 12 supernova remnants, nine of which have $\tau_{c} \in [2.5, 16.5]$~kyr. No evidence of a CW signal was found. The reported strain limits are slightly above $1\times 10^{-25}$ at a 90$\%$ confidence interval level and comparable to the limit obtained here. Papa~\textit{et al.}~\cite{Papa_2020} used O1 and O2 data to search for CW signals from the central, compact objects associated with three young supernova remnants; Cassiopeia A (Cas~A), Vela Junior (Jr.) and G347.3\textendash0.5. The authors set 90$\%$ confidence limits of $h_0^{90\%} = 1.2\times 10^{-25},\ 9.3\times 10^{-26}\ $ and $\ 9.0\times 10^{-26}$ for Cas~A, Vela~Junior (Jr.), and G347.3-0.5 near 185~Hz, respectively. The limits for Vela Jr. and G347.3-0.5 are $\sim 1.3$~times better than the constraint obtained here, while the Cas~A limit is comparable. Following this study, Ming \textit{et al.}~\cite{Ming_2022} improved the constraints on the 90$\%$ upper limit for G347.3-0.5 ($\tau_{c} = 1.6$~kyr) to $h_0^{90\%} \sim 7.5\times 10^{-26}$, which is 1.7~times better than this search, albeit for a different target~\cite{Ming_2022}. However, none of the above analyses track stochastic spin wandering, unlike the HMM approach in this paper. Studies with data from future observation runs and better analysis methods will further extend the sensitivity of CW searches and increase the chances of detection.\\

\begin{acknowledgments}
The authors would like to express their gratitude toward Ling Sun, Meg Millhouse, Hannah Middleton, Patrick Meyers, and Julian Carlin for numerous discussions and their ongoing support throughout this project. This research is supported by the Australian Research Council Centre of Excellence for Gravitational Wave Discovery (OzGrav) with Project No. CE170100004. This work used computational resources of the OzSTAR national facility at Swinburne University of Technology. OzSTAR is funded by Swinburne University of Technology and the National Collaborative Research Infrastructure Strategy (NCRIS). This research has made use of data, software and/or web tools obtained from the Gravitational Wave Open Science Center \cite{Collaboration_2019}, a service of LIGO Laboratory, the LIGO Scientific Collaboration and the Virgo Collaboration.
\end{acknowledgments}

\appendix
\section{EXPRESSION FOR $T_{\mathrm{coh}}$}~\label{app:t_coh}
In this search, we require the change in CW signal frequency due to the secular spin-down of the pulsar over [$t,t+T_{\rm{coh}}$] to satisfy
\begin{equation}
    \left| \int_t^{t + T_{\rm{coh}}} dt' \dot{f_0}(t') \right| < \Delta f_{\rm{coh}} \label{SD_relation1}
\end{equation}
for $ 0 < t < T_{\rm{obs}}$ \cite{Sun_2018}. Here, $T_{\rm{coh}}$ denotes the coherence timescale, $T_{\rm{obs}}$ denotes the total observation period and $\Delta{f_{\rm{coh}}}$ is the frequency resolution of the search, related to $T_{\rm{coh}}$ via 
\begin{equation}
    \Delta f_{\rm{coh}} = \frac{1}{2T_{\rm{coh}}}. \label{SD_relation2}
\end{equation}
Even though the signal frequency $f_0$ may not be locked to the star's spin frequency $f_\ast$ \cite{Bennett_1991,Melatos_2012}, we can assume $\dot{f}_0 \approx \dot{f}_\ast$ to a good approximation. Combining relation (\ref{SD_relation1}) with (\ref{SD_relation2}), we arrive at the following equality:
\begin{equation}
    |\dot{f}_*|T_{\rm{coh}} \approx \Delta f_{\rm{coh}} = \frac{1}{2T_{\rm{coh}}}.
\end{equation}
By solving for $T_{\rm{coh}}$, one obtains $T_{\rm{coh}}\leq (2|\dot{f}_*|)^{-1/2}$.

\section{VETOES\label{app:vetoes}}
The non-Gaussian nature of interferometer noise can cause outliers with detection statistic above the threshold. To remove such artifacts, all candidates with $\mathcal{L}>\mathcal{L}_{\rm{th}}$ are passed through a four-step veto procedure. Below, we briefly describe these vetoes and their rejection criteria. 
\begin{enumerate}
\item Instrumental noise lines.~\label{sub:vetos_1}A vast majority of the terrestrial candidates are identified and rejected using the list of persistent instrumental lines~\cite{Covas_2018}. These lines are identified during the detector characterization process and originate as resonant modes of the suspension system, external environmental causes and interference from the equipment around the detector. We veto all candidates for which [$f_0 -\delta f_0,f_0 +\delta f_0$] intersects with a known instrumental line. Here, $f_0$ is the frequency of the path at $t=0$ and $\delta f_0$ is the frequency spread due to the Doppler shift correction applied by the $\mathcal{F}$~statistic, i.e., $\pm\delta f_0 \approx 10^{-4}f_0$~\cite{Jones_2022}. 
    
    \item Single interferometer veto.~\label{sub:vetos_2}An astrophysical signal should be present in data from all detectors and have a better signal-to-noise ratio in the detector with higher sensitivity. Strong noise artifacts that are only present in one detector can also produce candidates with $\mathcal{L}>\mathcal{L}_{\rm{th}}$ in the dual-interferometer search. To separate local noise from astrophysical signals, we repeat the analysis in the Hanford and Livingston detectors individually. A candidate is vetoed if it satisfies two criteria: (1)~One of the single interferometer searches yields $\mathcal{L}\geq\mathcal{L}_{\cup}$, where $\mathcal{L}_{\cup}$ denotes the log-likelihood from the dual-interferometer search, while the other interferometer yields $\mathcal{L}<\mathcal{L}_{\cup}$, and (2)~the Viterbi path from the interferometer with $\mathcal{L}\geq\mathcal{L}_{\cup}$ intersects the original path.
    
    \item Doppler modulation (DM) veto.~\label{sub:vetos_3}This veto was first introduced in Ref.~\cite{Sylvia_2017} and further studied in Refs.~\cite{Jones_2021,Strang_2021,Jones_2022}. It involves turning off the DM correction, which accounts for the Doppler shift due to Earth's motion around the Sun, and recomputing the log-likelihood~\cite{Sun_2018,Jones_2021}. By default, the $\mathcal{F}$~statistic applies this correction to the data depending on the source's sky location. It boosts the significance of a true astrophysical signal while spreading a terrestrial signal over several bins, thus reducing its significance. Therefore, we veto a candidate if the DM-off analysis yields $\mathcal{L}_{\rm{DM-off}} \geq \mathcal{L}_{\cup}$ and a new Viterbi path which intersects the band [$f_0-\delta f_0, f_0+\delta f_0$], where $f_0$ denotes the frequency of the candidate at $t=0$. An injection study is used to test the validity of the DM veto in the search configuration presented in this paper. The results are summarized in Appendix~\ref{app:dm_veto_test}.
    
    \item Off-target veto.~\label{sub:vetos_4}
    First introduced in Ref.~\cite{Middleton_2020} and further studied in Refs.~\cite{Jones_2021,Strang_2021,Jones_2022}, this veto involves shifting the sky position away from the true source's location. The sky offset is related to the length of $T_{\rm{coh}}$ such that as $T_{\rm{coh}}$ increases, the offset decreases. An astrophysical candidate should yield the highest detection statistic at the source's sky position~\cite{Isi_2020}. On the contrary, a noise artifact will remain consistently above $\mathcal{L}_{\rm{th}}$ regardless of the offset. For this study, we adapt the sky offset from Ref.~\cite{Beniwal_2021}, as the coherence times are comparable. This involves shifting the right ascension by $\delta_{\rm{RA}} =3~\rm{h}$ and declination by $\delta_{\rm{DEC}} =10~\rm{min}$. We veto the candidate if the off-target search yields $\mathcal{L}_{\rm{off}} \geq \mathcal{L}_{\cup}$ and returns a new Viterbi path intersecting the band $[f_0 - \delta f_0,f_0 +\delta f_0]$ from the dual-interferometer, DM-on search.

\end{enumerate}

\section{DOPPLER MODULATION VETO~\label{app:dm_veto_test}}
We use software injections to test the effectiveness of the DM veto in separating an astrophysical signal from a terrestrial one. We start by finding a threshold log-likelihood ($\mathcal{L}_{\rm{th}}$) to give a desired false alarm probability of $\alpha_f = 1\%$ in the 100\textendash102~Hz sub-band. The procedure used here is identical to the one described in Sec.~\ref{subsec:threshold}. We generate 500 Gaussian noise-only realizations in the chosen sub-band and set the detector ASD [$S_h(f)^{1/2}$] to match the band-averaged one-sided noise ASD of the O2 data (see Table~\ref{tab:dm_off_injection}). We search the data using a combination of the $\mathcal{F}$~statistic and the Viterbi algorithm. The distribution of maximum log-likelihoods is then used to find the 99th percentile $\mathcal{L_{\rm{th}}}$ corresponding to $\alpha_f = 1\%$.\\

Here, we test the validity of the DM veto using injections into Gaussian noise in two regimes: the strong signal regime, where the injected signals are easily detectable (i.e., $\mathcal{L}>>\mathcal{L}_{\rm{th}}$), and the weak signal regime, where the signals are marginally above the detection threshold (i.e., $\mathcal{L}>\mathcal{L}_{\rm{th}}$). The Gaussian thresholds for the two cases are stated in Table~\ref{tab:dm_off_search}.\\

Once the thresholds have been established, we inject synthetic signals into white Gaussian noise. The injection parameters are outlined in Table~\ref{tab:dm_off_injection}. We generate 100 realisations for Hanford and Livingston detectors using the Makefakedata$\_$v4 tool in \texttt{LALSuite}~\cite{LALapps_2018}. The data are then searched using a combination of the $\mathcal{F}$~statistic and Viterbi algorithm. We summarise the parameters for this study in Table~\ref{tab:dm_off_search}. These are identical to ones used in the original search. Each candidate with $\mathcal{L}\geq \mathcal{L}_{\rm{th}}$ is searched again with DM turned off and rejected if it satisfies the criteria outlined in Appendix~\ref{app:vetoes}. \\

\begin{table}[h!]
    \renewcommand{\arraystretch}{1.2}
    \centering
    \caption{\small{Injection parameters used to create synthetic data analysed in the DM-veto study. rand($a,b$) denotes a uniformly distributed random number between $a$ and $b$.}}
    \begin{tabular}{l c c}
    \hline
    Parameter & Value & Unit \\
    \hline \hline
    Reference time & 1167545066 & ... \\
    RA & 14:27:56.7 & J2000 h:min:s \\
    DEC & \textendash60:52:14 & J2000 deg:min:s  \\
    Band & 99.9\textendash102.1 &~Hz \\
    $f_0$ & rand(100,102) &~Hz \\
    $\cos{\iota}$ & 0 & ... \\
    $h_0$ & (1.5 and 3) $\times10^{-25}$ & ... \\
    $S_h(f)^{1/2}$ &$5.6\times10^{-24}$ & $1/\sqrt{\rm{Hz}}$\\ \hline    
    \end{tabular}
    \label{tab:dm_off_injection}
\end{table}

\begin{table}[h!]
    \renewcommand{\arraystretch}{1.2}
    \centering
    \caption{\small{Search parameters used to test the validity of the DM veto.}}
    \begin{tabular}{l c c}
    \hline
    Parameter & Value & Unit \\
    \hline \hline
    Reference time & 1167545066 & ... \\ 
    RA & 14:27:56.7 & J2000 h:min:s \\
    DEC & \textendash60:52:14 & J2000 deg:min:s  \\
    Band & 100\textendash102 &~Hz \\
    $\Delta f$ & $1.9\times10^{-5}$ &~Hz \\ 
    $T_{\rm{coh}}$ & 7.5 & h  \\
    $T_{\rm{obs}}$ & 234 & days \\ 
    Threshold & 5429 and 5397 & ... \\ \hline    
    \end{tabular}
    \label{tab:dm_off_search}
\end{table}
 
\begin{figure*}
    \centering
    \begin{subfigure}[b]{\columnwidth}        
        \centering
        \caption{Strong signal regime}\label{subfig:strong_signal}       
        \includegraphics[trim=0cm 0cm 0cm 1.13cm, clip=true, width=\linewidth]{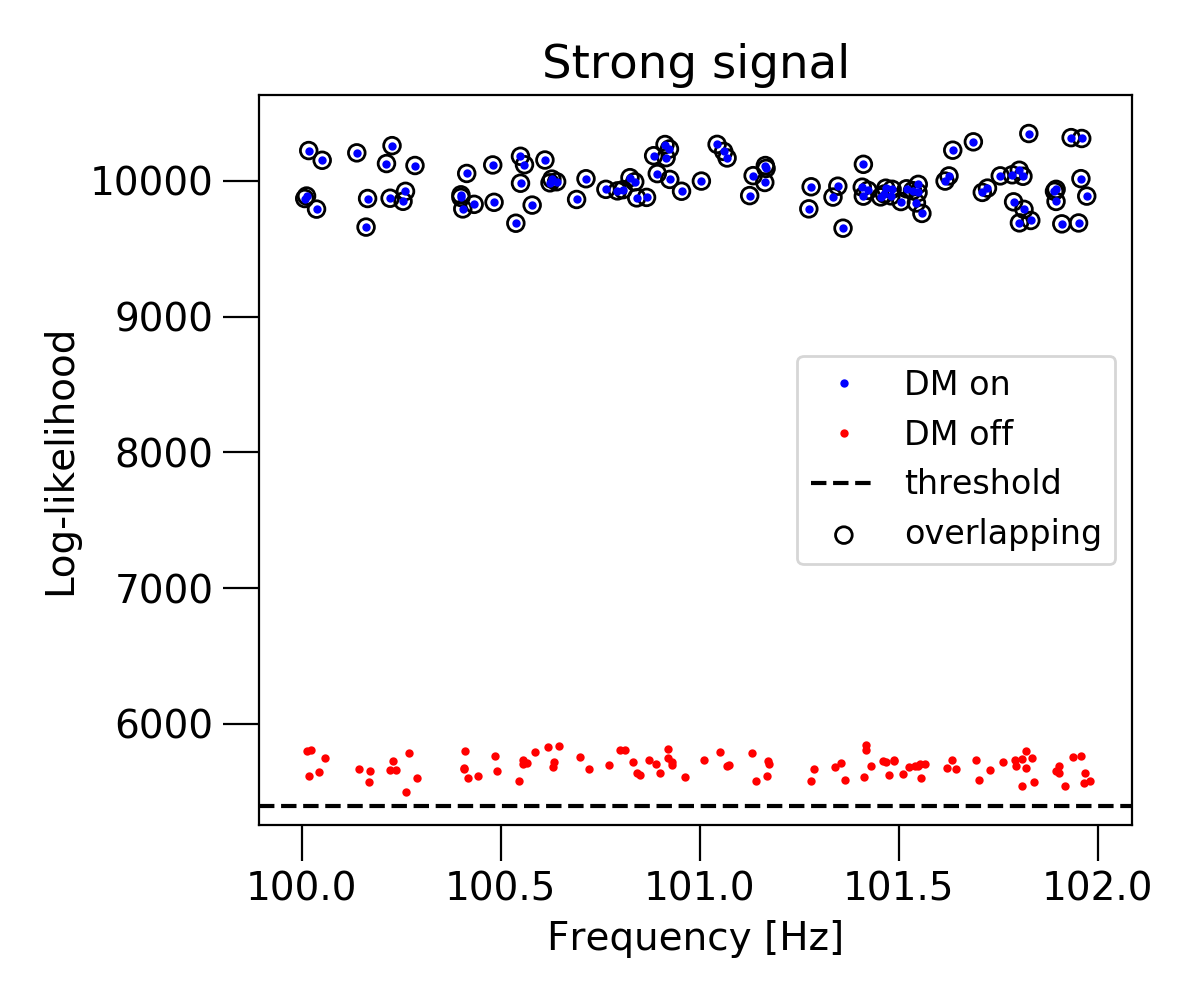}
    \end{subfigure}
    \hfill
    \begin{subfigure}[b]{\columnwidth}        
        \centering
        \caption{Weak signal regime}\label{subfig:weak_signal}
        \includegraphics[trim=0cm 0cm 0cm 1.13cm, clip=true, width=\linewidth]{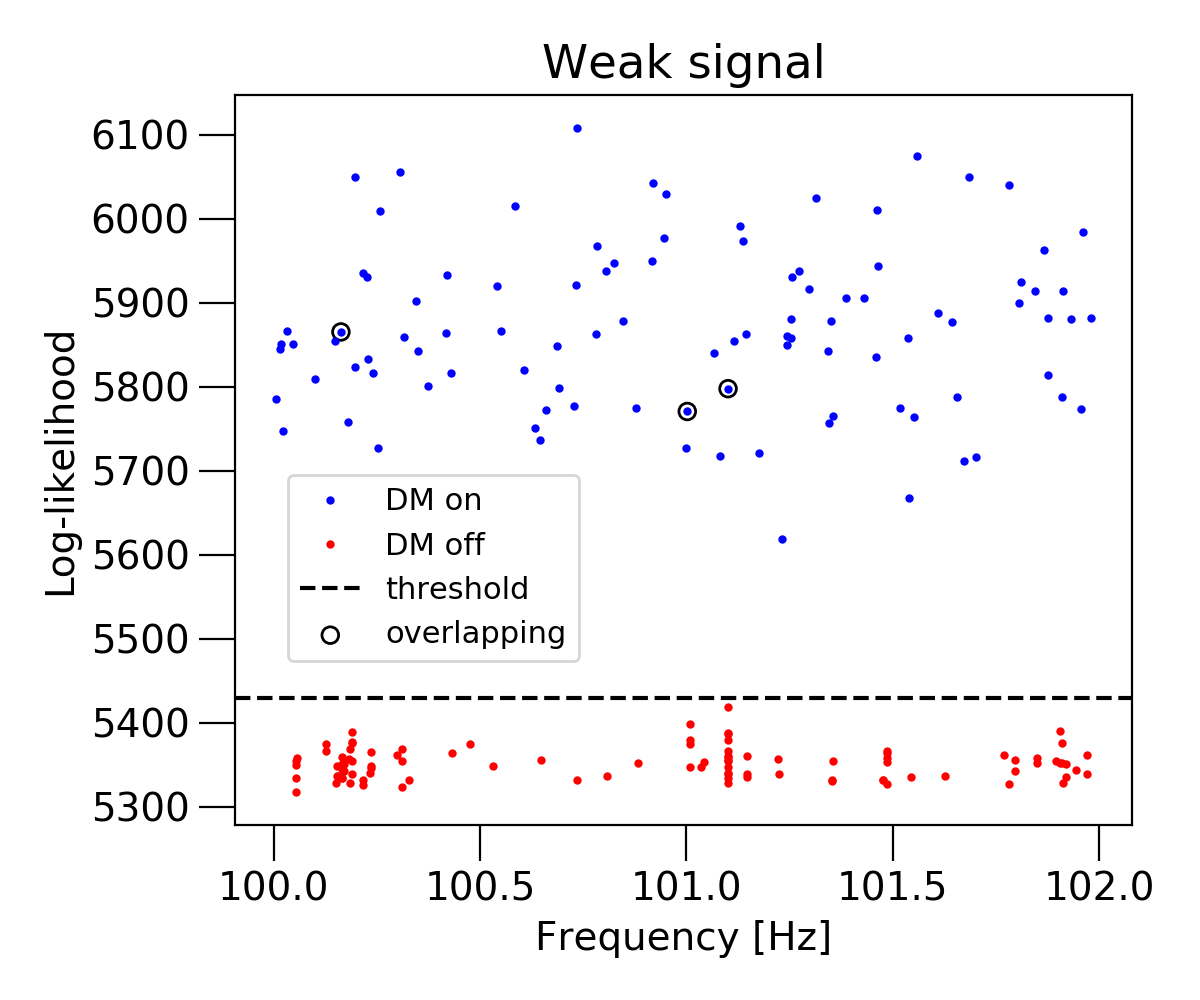}
    \end{subfigure}
    \caption{\small{Outcome of the Doppler modulation veto injection study, displayed as log-likelihood versus signal frequency $f_0$. (a) Results for the strong signal regime ($h_0=3\times 10^{-25}$). Blue and red dots show the results for a search with the Doppler modulation (DM) correction turned on and off, respectively. The circles indicate realisations that have overlapping paths between the two searches. (b) Weak injections ($h_0=1.5\times 10^{-25}$) in Gaussian noise, laid out as in (a). Note that all of the DM-off candidates are below the Gaussian threshold. Additionally, only 3/100 realisations return overlapping paths in the DM-off veto.}}
    \label{fig:DM_veto_study}
\end{figure*}

The results are shown in Fig.~\ref{fig:DM_veto_study}. We experiment with two different values of $h_0$ to investigate the validity of DM veto in the strong and weak signal regime. The blue and red dots in Fig.~\ref{fig:DM_veto_study} indicate the resultant log-likelihood with ($\mathcal{L}_{\cup}$) and without ($\mathcal{L}_{\rm{DM-off}}$) the DM correction, respectively. In the strong signal regime [Fig.~\ref{subfig:strong_signal}], all candidates in the DM-off search return Viterbi paths that intersect their DM-on counterpart in the band [$f_0-\delta f_0$, $f_0+\delta f_0$]. This is expected, as the injected signals have a sufficiently large signal-to-noise ratio to be easily separated from the background noise, even when the DM is turned off. However, all of these candidates have $\mathcal{L}_{\rm{DM-off}}\ll\mathcal{L}_{\cup}$, thus only satisfying one out of the two veto criteria. As a result, the injections pass the DM veto. Similarly, all candidates in the weak signal regime [Fig.~\ref{subfig:weak_signal}] return $\mathcal{L}_{\rm{DM-off}}<\mathcal{L}_{\rm{th}}<\mathcal{L}_{\cup}$ and thus pass the DM veto. This is true even if the analysis returns candidates with overlapping paths, as is the case for 3 out of 100 realisations. The lack of overlapping paths is expected, as the signals become indistinguishable from background noise at this strain sensitivity. These results suggest that we can safely apply the DM veto outlined in Appendix~\ref{app:vetoes} to this search. 

\newpage
\bibliography{references}
\end{document}